\documentclass[manuscript,screen,sigchi]{acmart}

\AtBeginDocument{%
  \providecommand\BibTeX{{%
    \normalfont B\kern-0.5em{\scshape i\kern-0.25em b}\kern-0.8em\TeX}}}

\setcopyright{acmcopyright}
\copyrightyear{2018}
\acmYear{2018}
\acmDOI{XXXXXXX.XXXXXXX}

\acmConference[SIGCHI '25]{April 2025}{Yokohama, Japan}

\acmSubmissionID{12345}

\usepackage{xcolor}
\usepackage{todonotes}
\usepackage{graphicx}
\usepackage{csquotes}

\usepackage[utf8]{inputenc} 
\usepackage[T1]{fontenc}    
\usepackage{hyperref}       
\usepackage{url}            
\usepackage{booktabs}       
\usepackage{amsfonts}       
\usepackage{nicefrac}       
\usepackage{microtype}      
\usepackage{lipsum}
\usepackage{fancyhdr}       
\usepackage{graphicx}       
\usepackage{caption}
\usepackage{subcaption}
\usepackage{wrapfig}

\begin{document}

\title[Toward a Human-Centered AI-assisted Colonoscopy System]{Toward a Human-Centered \\ AI-assisted Colonoscopy System in Australia}


\author{Hsiang-Ting Chen}
\affiliation{%
  \institution{University of Adelaide}
  \city{Adelaide}
  \country{Australia}
}
\email{tim.chen@adelaide.edu.au}

\author{Yuan Zhang}
\affiliation{%
  \institution{University of Adelaide}
  \city{Adelaide}
  \country{Australia}
}
\email{yuan.zhang01@adelaide.edu.au}

\author{Gustavo Carneiro}
\affiliation{%
  \institution{University of Surrey}
  \city{Surrey}
  \country{United Kingdom}
}
\email{g.carneiro@surrey.ac.uk}

\author{Rajvinder Singh}
\affiliation{%
  \institution{Lyell McEwin Hospital}
  \city{Adelaide}
  \country{Australia}
}
\email{rajvinder.singh@sa.gov.au}

\begin{abstract}


While AI-assisted colonoscopy promises improved colorectal cancer screening, its success relies on effective integration into clinical practice, not just algorithmic accuracy.  This paper, based on an Australian field study (observations and gastroenterologist interviews), highlights a critical disconnect: current development prioritizes machine learning model performance, overlooking essential aspects of user interface design, workflow integration, and overall user experience.  Industry interactions reveal a similar emphasis on data and algorithms. To realize AI's full potential, the HCI community must champion user-centered design, ensuring these systems are usable, support endoscopist expertise, and enhance patient outcomes.

\end{abstract}




\maketitle



\section{Introduction}

Bowel cancer is the second most common and deadliest cancer in Australia \cite{cancerreport}. The National Bowel Cancer Screening Program (NBCSP) recommends colonoscopy for high-risk individuals (e.g., family history, age) or those with a positive FOBT. Colonoscopy aims to detect and remove precancerous polyps, reducing cancer risk \cite{cancerreport}.

However, colonoscopy's accuracy is limited by factors like endoscopist skill, bowel preparation, and fatigue, leading to missed or misdiagnosed polyps. While a high adenoma detection rate (ADR) is inversely correlated with cancer risk \cite{Kaminski2017-wp}, and visual assessment guides polyp management, challenging cases necessitate histopathology, incurring risks and costs. Meta-analyses document high missed polyp (22-27\% \cite{Zhao2019-wg, Van_Rijn2006-bq}) and missed cancer rates (up to 8

To address these limitations, AI-powered polyp detection and classification have advanced rapidly.  Machine- and deep-learning models show impressive results in retrospective and prospective studies \cite{Hassan2021-survey}, and randomized controlled trials confirm CADe system effectiveness. A meta-analysis projects a ~52\% ADR increase \cite{shah2023_Meta}, significantly improving cancer detection. Consequently, manufacturers are deploying commercial AI products offering real-time visual assistance.
However, despite these promising results, integrating AI into clinical practice raises concerns. Early-adopter endoscopists report AI fatigue from information overload~\cite{Mori2021-next} and express concerns about the lack of AI model transparency, potentially leading to incorrect recommendations and patient harm.

This paper shares our experience developing AI-assisted colonoscopy systems in Australia over the past two years, targeting the HCI community. We have developed multiple machine learning models for both upper and lower gastrointestinal (GI)~\cite{butler2022defense,tan2023enhancing,tan2024exploring} and conducted a field study involving semi-structured interviews with three gastroenterologists of varying experience levels at a local hospital. This research aimed to understand the specific needs of these clinicians. AI-assisted colonoscopy systems occupy a unique position: while the underlying AI algorithms are sufficiently robust to gain regulatory approval for clinical use and assist in high-stakes decision-making, the usability of these systems is often considered poor by clinicians \cite{Mori2021-next}. We believe the field of AI-assisted colonoscopy urgently requires the attention of the HCI community. Incorporating a human-centered design approach can significantly increase clinician adoption and acceptance, ultimately leading to better patient outcomes both in Australia and globally.

\section{Background and Related Work}



Colorectal cancer (CRC), a common malignancy of the digestive tract, develops in the colon. In Australia (2019), it was the second most commonly diagnosed cancer and the deadliest, with 16,398 new cases and 5,597 deaths \cite{PanJennifer2019Oosa}. CRC typically originates as polyps – flat, raised, stalked, or carpet-like growths along the colonic lumen. 


\begin{figure} [t]
    \centering
    \includegraphics[width=\linewidth]{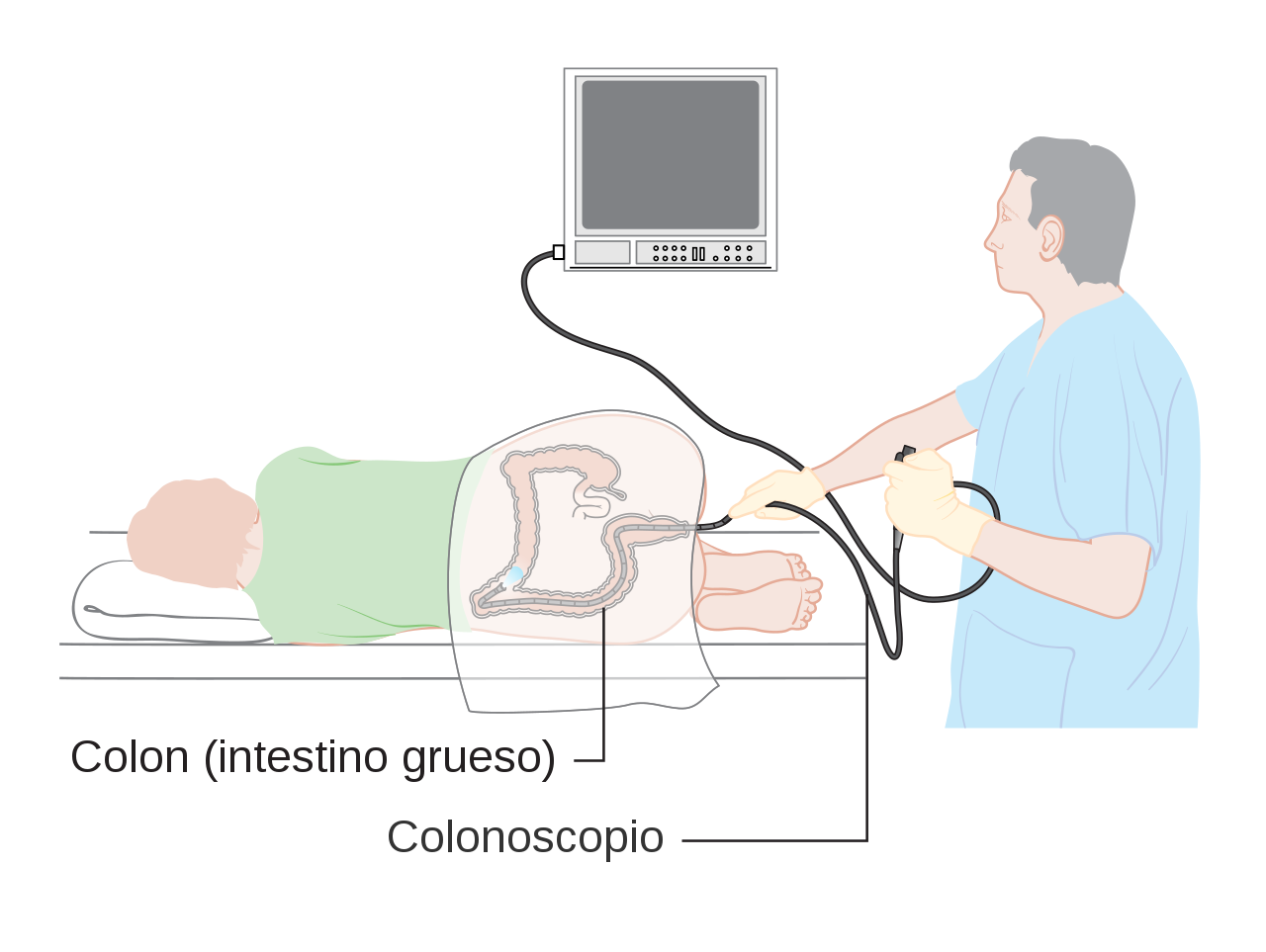}
    \caption{The colonoscopy procedure diagram.}
    \label{fig:colonoscopy}
\end{figure}

Colonoscopy is the most effective method for reducing colorectal cancer incidence and mortality \cite{cancerreport}. During this 30-60 minute procedure, a fiber-optic camera on a flexible tube (Figure~\ref{fig:colonoscopy}) examines the colon, displaying video on an external screen for visual diagnosis. Endoscopists visually assess discovered polyps as hyperplastic, adenomatous, or malignant, determining subsequent management. If a reliable optical diagnosis is impossible, histopathological assessment of biopsied or resected polyps is required, a practice carrying risks, inefficiencies, and high costs.


Polyp detection and classification accuracy depend on endoscopist factors (experience, fatigue), operational factors (bowel preparation, withdrawal time), and lesion characteristics \cite{murphy2020correlation}. While new endoscopic technologies like high-definition colonoscopy, narrow-band imaging (NBI), and intelligent color enhancement have improved visualization \cite{kim2017miss,subramanian2011high,nagorni2012narrow}, their impact on polyp detection remains controversial due to the need for more experienced endoscopists, longer procedure times, and increased costs.

\subsection{Computer-Aided Detection of Polyps}

Early research explored computer-aided colonoscopy, analyzing video alongside the endoscopist \cite{karkanis2001evaluation,tjoa2003feature,zheng2005fusion}. Deep learning now dominates CADe, outperforming traditional methods in competitions like MICCAI 2015 \cite{bernal2017comparative}.
For a CADe overview, see Kumar et al. \cite{SurveyPMC2023} and the "Awesome Polyp Segmentation" repository \cite{AwesomePolypSegmentation2024}. Recent work focuses on real-time, efficient deep learning, primarily CNNs. Studies use datasets like Kvasir-SEG, with new architectures (e.g., ColonSegNet \cite{Tuba2021}) balancing accuracy and speed. Lightweight models like Enhanced NanoNet \cite{Khan2024} and Jha et al.'s \cite{Jha2024} improve segmentation with minimal computation, achieving high Dice scores\cite{DeFrame2022}. 

Few works investigated colonoscopy using HCI methodologies. Van Berkel et al. \cite{Van_Berkel2021-marker} evaluated seven visual markers for polyp detection and concluded that the clinicians preferred blue wide bounding circles. Notably, the experiment did not assess the position map design from the CAD EYE system (Figure~\ref{fig:commercial} b). AI-assisted colonoscopy has also been discussed as a case study under the theme of continuous human-AI interaction \cite{Wintersberger2022-continuous,Van_Berkel2022-continuous}.

 
\subsection{Commercial AI-assisted Colonoscopy Systems}
\begin{figure}
    \centering
    \includegraphics[width=0.9\linewidth]{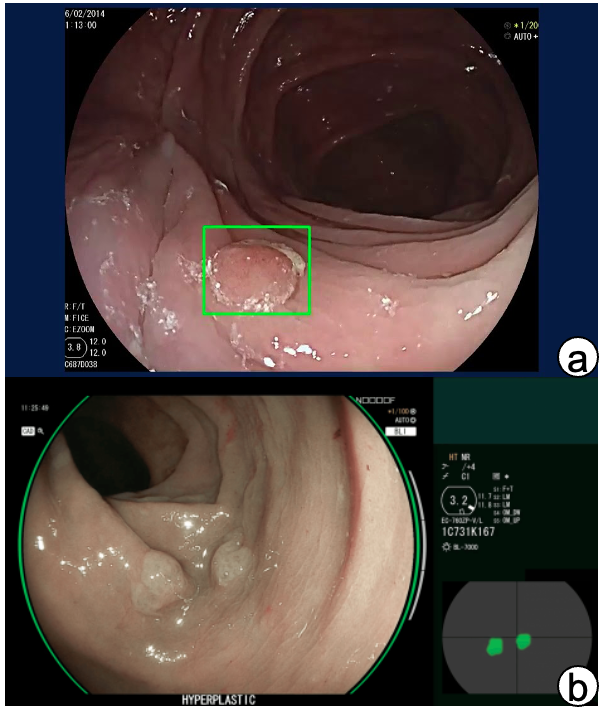}
    \caption{Commercial AI-assisted colonoscopy systems}    
    \label{fig:commercial}   
\end{figure}

Driven by impressive AI model results, several commercial AI-assisted colonoscopy systems have emerged, including Medtronic \footnote{\url{https://www.medtronic.com/covidien/en-us/products/gastrointestinal-artificial-intelligence/gi-genius-intelligent-endoscopy.html}}, the \textit{EndoBRAIN} by Cybernet \footnote{\url{https://www.cybernet.co.jp/medical-imaging/products/endobrain/}} , \textit{AI4GI} by Satisfai \footnote{\url{https://satisfai.health/gi-cancer-treatments}}, and the \textit{CAD EYE} from Fujifilm \footnote{\url{https://fujifilm-endoscopy.com/cadeye}}
These systems integrate as extension modules with existing Olympus or Fujifilm endoscopes, taking video input and performing real-time polyp detection using CNN-based algorithms, often with YOLO-based tracking \cite{liu2020deep}. CADe assistance varies (Figure~\ref{fig:commercial}), but all overlay information directly onto the endoscope video stream. GI-Genius and AI4GI use bounding boxes around detected polyps (Figure~\ref{fig:commercial}a). EndoBRAIN avoids visual obstruction by using alert sounds and yellow corner highlights. CAD EYE offers multiple options: bounding boxes, a visual assistance circle, and a separate position map (Figure~\ref{fig:commercial}b). 
Furthermore, EndoBRAIN-plus and CAD EYE Characterisation provide real-time polyp classification (non-neoplastic/adenoma/invasive cancer for EndoBRAIN-plus; hyperplastic/neoplastic for CAD EYE), displaying results textually alongside the video.

\section{Field Study and Project Experience}

We conducted a field study at Lyell McEwin Hospital, South Australia, to understand colonoscopy procedures and the potential for AI assistance. A team (one UX designer and two AI researchers) observed a 30-minute colonoscopy performed by a senior gastroenterologist, assisted by a junior gastroenterologist, nurse, and anesthetist, using an Olympus X1 system with a 4K display. This procedure did not include AI assistance, which remains uncommon in Australia. Following the observation, we conducted separate semi-structured interviews with the senior gastroenterologist and two junior gastroenterologists (all specialist registrars familiar with CADe systems). The senior gastroenterologist had experience with Fujifilm's CAD EYE system. Interview questions are listed in Appendix~\ref{sec:questions}. Subsequently, we held several follow-up meetings with both the gastroenterologists and potential commercial partners to discuss system sketches and the potential technology transfer of the team's AI models \cite{butler2022defense,tan2024exploring}.

\section{Result and Discussion}
This section reports our initial findings from the field study and insights gained during subsequent discussions.

\begin{figure*}[ht!]
    \centering
    \includegraphics[width=0.9\linewidth]{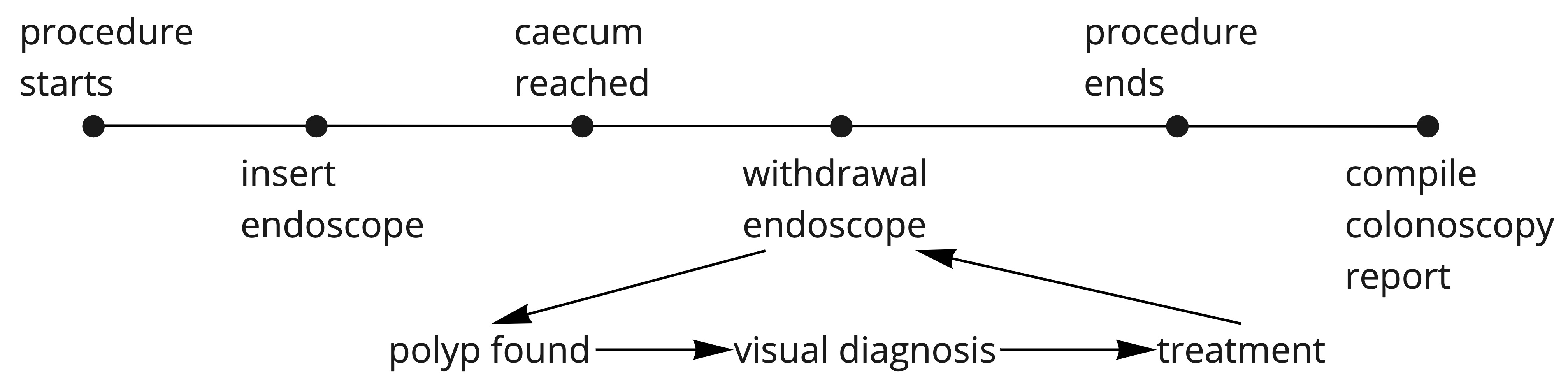}
    \caption{The colonoscopy procedure}
    \label{fig:colon_proc}
\end{figure*}
\subsection{Colonoscopy Procedure}
We summarise the colonoscopy procedure learned from the direct observation and the interview in Figure~\ref{fig:colon_proc}. The colonoscopy procedure begins with the endoscopist inserting the endoscope into the colon. Through the visual inspection, the endoscopist needs to continue the intubation until the endoscope reaches the caecum,i.e. the tip of the endoscope touches the appendiceal orifice. After reaching the caecum, the endoscopist slowly withdraws the endoscope and examines the colon's inner lining. When the endoscopist found a polyp, the endoscopist would maneuver the endoscope for a close-up view of the polyp and takes a screenshot for documentation. The endoscopist then classifies the polyp based on the visual inspection and decides a treatment, such as no intervention or removal and retrieval for pathology analysis. The polyp detection, diagnosis, and treatment process continues until the endoscope is wholly withdrawn. After the procedure is completed, the endoscopists then sit in front of a computer and compile a comprehensive colonoscopy report based on the captured photos during the colonoscopy. The report includes essential information such as  assessments of patient risk and comorbidity, the findings of polyps and interventions, and the follow-up plan.

\subsection{Polyp Detection and Classification (Q1, Q2, Q3, Q6)}
Gastroenterologists identified polyp shape, location, and type as key factors influencing detection difficulty. Small, flat, or sessile serrated adenomas are challenging. Polyps behind folds, in blind areas (medical wall of ascending colon, hepatic flexure, between appendiceal orifice and ileocecal valve, rectosigmoid junction), far from the endoscope, or at the periphery of view are also harder to detect.

A case difficult for an endoscopist might be simple for AI, and vice-versa.  For example, a tiny polyp, hard to see for a human, might be clear to an AI in a 4K image. Conversely, AI might be confused by lighting, feces, or water, while a trained endoscopist would not. Further research should investigate context-aware AI assistance.

Interviewees emphasized the importance of combined polyp detection and classification.  Endoscopic societies advocate for >90\% accuracy in optical diagnosis to enable cost-effective strategies (e.g., ignoring small, benign hyperplastic polyps; discarding resected adenomas without histopathology) \cite{rees2017narrow}.  General endoscopists often struggle to achieve this, necessitating removal and histopathological diagnosis of all polyps.  An AI system capable of expert-level diagnosis based on multiple features could save significant medical resources.

\subsection{Quality Assurance (Q4, Q5, Q7)}

Beyond polyp detection and classification, interviewees suggested using AI to audit colonoscopy procedures. Factors like poor withdrawal technique, short withdrawal time, inadequate bowel preparation, and failure to intubate the cecum can lead to missed polyps, regardless of AI assistance. An AI could monitor withdrawal time (optimally >7 minutes) by recognizing cecal landmarks.  It could also gather contextual data (polyp count, removal times, locations) to provide feedback and ensure adherence to standards.  Furthermore, interviewees desired AI automation of routine tasks: capturing polyp screenshots at varying distances/angles, estimating polyp size (correlating with cancer risk), and, ultimately, generating comprehensive procedure reports using natural language. While research exists on medical information extraction from reports \cite{wang2018clinical} and automated pathology report generation \cite{raju2015natural}, colonoscopy procedure report generation remains largely unexplored.

\subsection{More Challenges in AI-Assisted Colonoscopy}

Randomized controlled trials show AI-assisted colonoscopy can increase ADR to 36.6\% (vs. 25.2\% without AI). A ~4\% ADR increase reduces colorectal cancer rates by ~1\% in follow-up exams \cite{Kaminski2017-wp}, suggesting AI use might become preferred or even legally required.  However, current legal frameworks hold domain experts accountable for their decisions and patient safety \cite{Schneeberger2020-legal}. Thus, AI systems must not only aid in polyp detection/classification but also assist endoscopists in explaining, justifying, and taking responsibility for their decisions.

Conversely, over-reliance on AI could lead to negative outcomes, such as endoscopists relaxing their polyp search.  Superior AI performance might also reduce junior doctors' confidence and lead to deskilling \cite{sinagra2021deskill}. Future studies should explore the effects of prolonged AI assistance and optimal training methods.

Australia faces unique challenges due to urban/rural disparities in resources and demographics.  AI model training data, primarily from urban teaching hospitals, may introduce bias, affecting accuracy.  Rural hospitals with older, lower-resolution equipment may also present challenges to AI model validity. Similar to HCI research in developing countries \cite{wang2021-dev,karusala2020-dev} and AI use in rural healthcare \cite{Wang2021-rural}, addressing data and demographic biases is crucial for equitable AI-enabled colonoscopy in Australia.

\subsection{The Need for Integrated Design in AI-Assisted Colonoscopy}
Gastroenterologists we interviewed primarily focused on the machine learning model's performance, giving less consideration to user interface design. Industry partners exhibited a similar pattern: large healthcare companies prioritized data and annotation acquisition, while start-ups planned to address user experience internally, despite their interest in the AI models. This suggests a tendency to treat the AI algorithm and the user interface as separate entities, rather than as integral components of a single system. This separation, while understandable from a purely technical perspective, neglects the crucial interplay between algorithmic accuracy and user interaction, a core principle of HCI~\cite{Norman2013}. 
The effectiveness of AI-assisted colonoscopy depends not just on what the AI can do, but on how endoscopists can effectively interact with and interpret its output. The HCI community needs to advocate for a more integrated approach, where user needs and workflow considerations are incorporated from the earliest stages of AI development.  Encouragingly, Australian funding agencies are increasingly valuing co-design and user-centered approaches in medical technology, indicating a potential shift towards more holistic AI development.

\section{Conclusion}
AI-assisted colonoscopy holds significant promise for improving colorectal cancer detection and management, as evidenced by advances in polyp detection, classification, and procedural documentation. However, our field study and analysis reveal critical challenges that must be addressed to realize this potential fully. These include concerns about the impact of prolonged AI use on endoscopist skill, the need for robust quality assurance measures, and the potential for data bias to exacerbate existing healthcare disparities, particularly in rural Australia. 
Moreover, the prevailing emphasis on algorithmic accuracy over user experience underscores the vital role of HCI in ensuring that these systems are designed for seamless integration into clinical workflows and are truly supportive of endoscopist expertise. 
A collaborative, human-centered approach, bringing together HCI researchers, AI researchers, clinicians, and industry, is essential to navigating these challenges and maximizing the benefits of AI for the society.


\bibliographystyle{ACM-Reference-Format}
\bibliography{reference}

\appendix
\section{Questions for the Semi-structured Interview}
\label{sec:questions}
\subsection*{AI-Assisted Colonoscopy - During Procedure}
\begin{itemize}
    \item Q1. What kind of polyps are most challenging for an endoscopist to detect or characterize?
    \item Q2. What kind of information from AI can assist you in such challenging situations?
    \item Q3. What are the common mistakes you have found with junior endoscopists? What type of assistance would a junior endoscopist often need during the procedure?
    \item Q4. During the procedure, the endoscopist constantly engages in multiple tasks such as searching for polyps, cleaning up the colon with water, characterizing polyps, and removing polyps. Are there other frequent tasks that you think AI can assist?
    \item Q5. AI can automatically analyze and record different events in the procedure. For example, an AI can automatically record the number of polyps detected, the amount of time for each polyp removal, or automatically take photos of the polyps. Would such functions be useful to you? Are there anything that you hope AI can automate for you?
    \item Q6. What is the last thing you want AI to do during the procedure?
\end{itemize}

\subsection*{AI-Assisted Colonoscopy - After-Procedure}
\begin{itemize}
    \item Q7. After the procedure, the endoscopist needs to compile and annotate the procedure record. What are the common difficulties you encounter at this step? Is there anything that you wish could be automated?
\end{itemize}



\end{document}